\documentclass[]{spie}  

\usepackage{amsmath,amsfonts,amssymb}
\usepackage{graphicx}
\usepackage[colorlinks=true, allcolors=blue]{hyperref}

\usepackage{xcolor}
\usepackage{colortbl}
\definecolor{c0}{HTML}{213652}
\definecolor{c1}{HTML}{3A5D8B}
\definecolor{c2}{HTML}{84AABF}
\definecolor{c3}{HTML}{F2C380}
\definecolor{c4}{HTML}{F29A79}

\title{The Roman Coronagraph Community Participation Program:
Observation planning and data reduction for polarimetric mode}

\author[a]{Ramya M Anche}
\author[b,c]{Toshiyuki Mizuki}
\author[a]{Justin Hom}
\author[d]{Alexis Lau}
\author[e]{Saanika Choudhary}
\author[f]{Jaren N. Ashcraft}
\author[g]{Clarissa Do O}
\author[h]{Tsutsumi Nagai}
\author[d]{Sophie Noiret}
\author[f]{Eric Shen}
\author[b,c]{Taichi Uyama}
\author[i]{Chen Xie}
\author[f]{Jingwen Zhang}
\author[j]{Vanessa P. Bailey}
\author[j]{Eric Cady}
\author[k]{Jessica Gersh-Range}
\author[i]{Julien H. Girard}
\author[n]{Guillermo Gonzalez}
\author[b,c]{John Livingston}
\author[j]{Bertrand Mennesson}
\author[f]{Maxwell A. Millar-Blanchaer}
\author[j]{Julia Milton}
\author[b,c]{Naoshi Murakami}
\author[e]{{Dmitry Savransky}}
\author[l]{Motohide Tamura}
\author[m]{Jason J. Wang}
\author[a]{Schuyler G. Wolff}
\author[j]{Marie Ygouf}

\affil[a]{Steward Observatory, University of Arizona, 933N Cherry Avenue, Tucson, Arizona, 85721, USA}
\affil[b]{National Astronomical Observatory of Japan, NINS, 2-21-1 Osawa, Mitaka, Tokyo 181-8588, Japan}
\affil[c]{Astrobiology Center, NINS, 2-21-1 Osawa, Mitaka, Tokyo 181-8588, Japan}
\affil[d]{Aix Marseille Univ, CNRS, CNES, LAM, Marseille, France}
\affil[e]{Sibley School of Mechanical and Aerospace Engineering, Cornell University, Ithaca, NY, 14853, USA}
\affil[f]{University of California, Santa Barbara, CA 93106, USA}
\affil[g]{Caltech, 1200 E. California Blvd., Pasadena, CA, 91125 USA}
\affil[h]{The Graduate Univ. for Advanced Studies, Shonan Village, Hayama, Kanagawa 240-0193, Japan}
\affil[i]{Space Telescope Science Institute, 3700 San Martin Drive, Baltimore, MD 21218, USA}
\affil[j]{Jet Propulsion Laboratory, California Institute of Technology, Pasadena, CA 91109, USA}
\affil[k]{DM Telescopes LLC, Raleigh, NC, USA}
\affil[l]{Department of Astronomy, Graduate School of Science, The University of Tokyo, 7-3-1, Hongo, Bunkyo-ku, Tokyo 113-0033, Japan}
\affil[m]{Northwestern University, 633 Clark Street Evanston, IL 60208, USA}
\affil[n]{Tellus1 Scientific, LLC, 8401 Whitesburg Dr SE, Unit 4662, Huntsville, AL 35802 USA}

\authorinfo{Further author information: (Send correspondence to R.M.A.)\\R.M.A.: E-mail: ramyaanche@arizona.edu}

\begin{document} 
\maketitle

\begin{abstract}
Reflected-light polarimetry of exoplanets constrains and resolves degeneracies in atmospheric properties, while polarized light observations of debris disks enable the characterization of dust-grain properties. The best-effort polarimetric mode of the Roman Coronagraph Instrument will be able to perform multi-wavelength observations of planetary systems using both the Hybrid Lyot Coronagraph (HLC) and the Shaped Pupil Coronagraph (SPC). This paper presents an overview of observation planning, simulations, and data reduction procedures for the polarimetric mode of the Roman Coronagraph. As an initial test of simulation and data reduction, a dataset of polarimetric observing sequences for the debris disk HD 172555 in HLC mode was generated using \texttt{corgisim} with estimated observation parameters, and data reduction was performed using \texttt{corgidrp}, incorporating all relevant noise factors and calibration products. Currently, mock calibration products are used in \texttt{corgidrp}; these will be replaced with simulated calibration products in future updates.

\end{abstract}


\section{Introduction}
One of the primary objectives of planetary astronomy is detecting and characterizing planets orbiting sun-like stars, with the aim of identifying potential biosignatures. Circumstellar debris disks, composed primarily of dust planetesimals, are critical for understanding ice lines, the composition of ice-giant and rocky planets, planet formation processes, and the overall architecture of planetary systems \cite{hughes2018debris,backman2004debris}. Despite the observation of scattered light from cold debris disks around approximately a dozen stars over the past two decades, accurately constraining the dust grain properties within these disks remains a significant challenge \cite{arriaga2020multiband}. Dust grain properties influence radiative transfer processes in complex and often degenerate ways, complicating efforts to disentangle and individually constrain these characteristics using simplistic dust models. Laboratory measurements indicate that assessing the linear polarization fraction provides a more effective diagnostic for constraining grain properties than total intensity observations alone \cite{munoz2021}. Consequently, comprehensive multiwavelength observations of both total intensity and polarization in circumstellar debris disks, utilizing next-generation space- and ground-based coronagraphic instruments, are essential for improving constraints on the scattering and geometrical properties of dust grains.

The Roman Coronagraph instrument on the Nancy Grace Roman Space Telescope, scheduled for launch readiness on 30 August 2026, is expected to provide unprecedented sensitivity to reflected light from exoplanets and circumstellar disks, achieving a planet-to-star contrast better than $10^{-7}$. Although the primary objective of the Roman Coronagraph (RC) is to satisfy the Technology Threshold Requirement (TTR5) in Band 1 ($\rm \lambda_c=575nm$) using the Hybrid Lyot Coronagraph (HLC)\cite{poberezhskiy2021roman,poberezhskiy2022roman}, the instrument is also capable of conducting multiwavelength total and polarization intensity observations as a best-effort mode. These observations are essential for advancing understanding of the architecture and composition of planetary systems. Upon achieving TTR5, the Roman coronagraph will demonstrate its performance through these best-effort modes and may further extend the mission to evaluate unsupported operational modes.

\begin{figure}[!ht]
    \centering
    \includegraphics[width=0.49\linewidth]{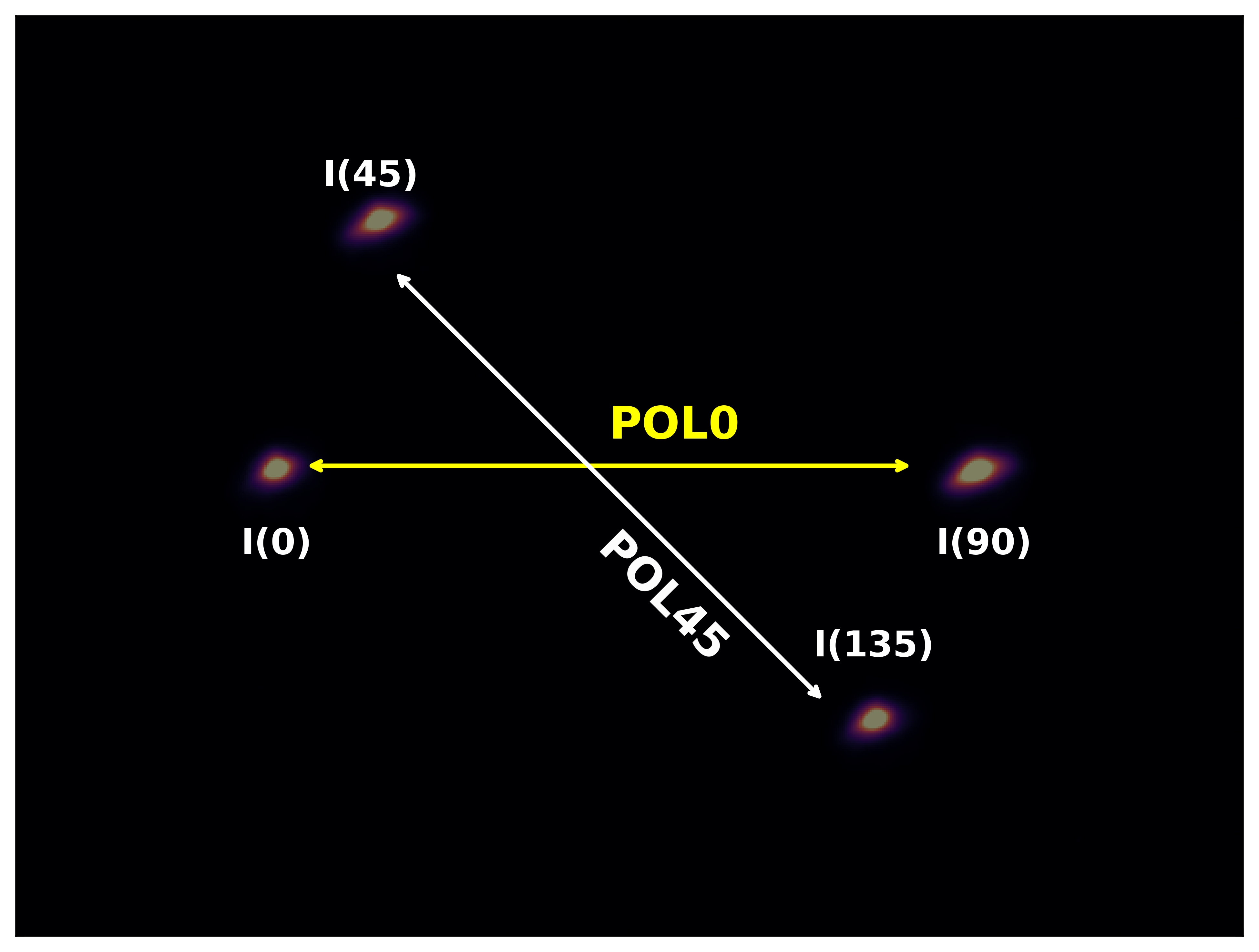}
    \includegraphics[width=0.49\linewidth]{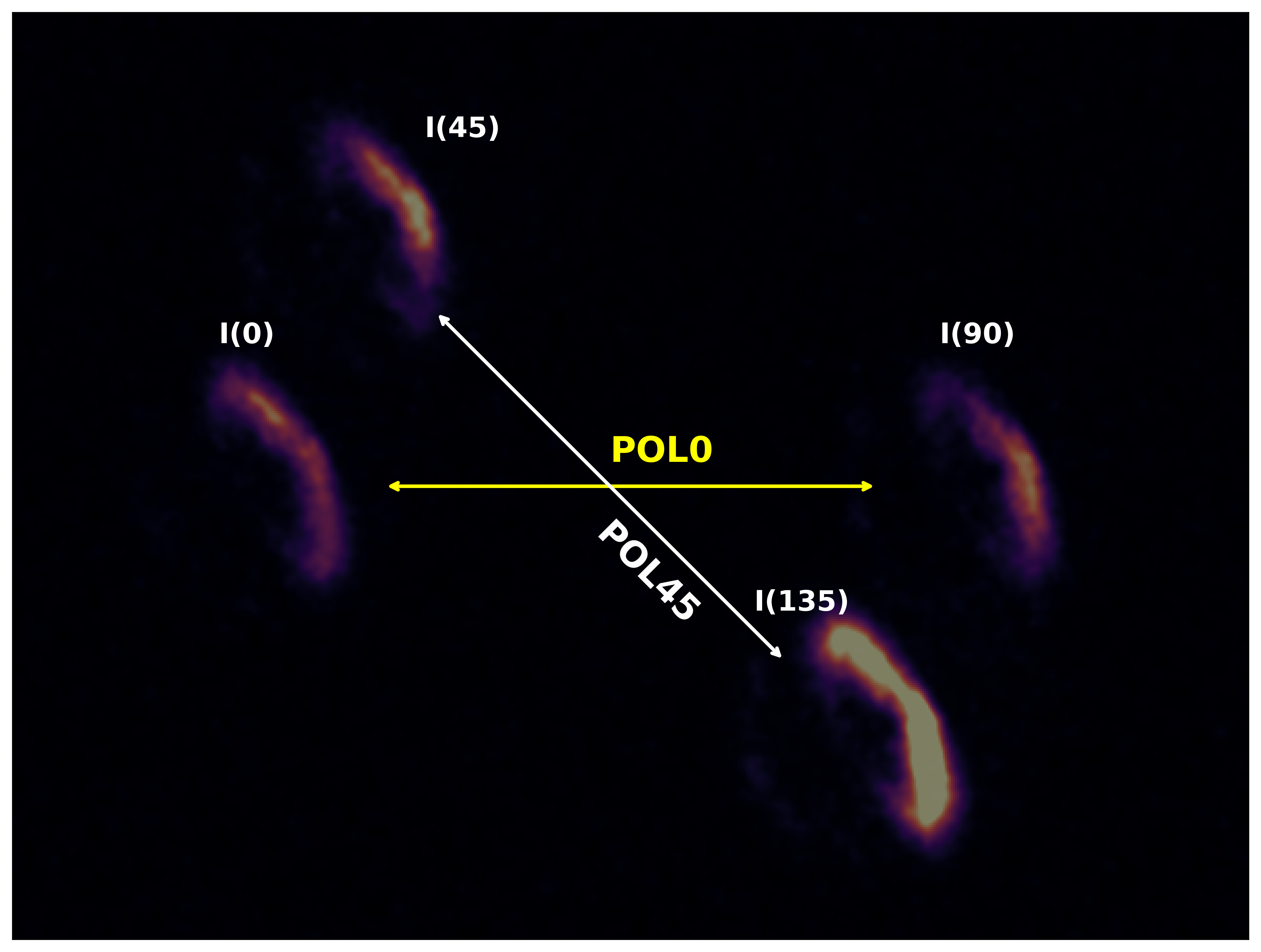}
    \caption{Roman Coronagraph Exoplanet Camera (EXCAM) image showing all the four orthogonal components for the HLC mode (left) and SPC mode (right)}
    \label{fig:excam_image}
\end{figure}

The polarimetric module of the Roman Coronagraph consists of two Wollaston prisms, POL0 and POL45, within the dispersion polarization alignment mechanism (DPAM), as illustrated in Figure 3 of Groff et al. 2025 \cite{groff2025spectroscopy}. These prisms split the incoming light into four orthogonal components: I0, I90, I45, and I135. Table \ref{tab:coronagraph-modes} presents the three best-effort polarimetric modes alongside their respective coronagraph and filter parameters.
The four orthogonal components on the detector called the exoplanetary systems camera (EXCAM) are shown for HLC (in band 1) and SPC (band 4) modes in Figure \ref{fig:excam_image}.
\begin{table}[!ht]
\begin{center}
\begin{tabular}{|l|l|l|l|l|l|l|} \hline
\rowcolor{c0}{\textcolor{white} {\textbf{Coronagraph}}} & {\textcolor{white} {\textbf{Config name}}} & {\textcolor{white} {\textbf{Band}}} & {\textcolor{white} {\textbf{Wavelength}}} & {\textcolor{white} {\textbf{Bandpass}}} & {\textcolor{white} {\textbf{IWA ($\lambda/D$)}}} & {\textcolor{white} {\textbf{OWA ($\lambda/D$)}}}       \\ \hline
\hline
\rowcolor{c4} HLC & Narrow FOV         & 1    & 575        & 10\%     & 3.0         & 9.7         \\ \hline 
\hline
\rowcolor{c3}SPC & Wide FOV    & 1    & 575        & 10\%     & 5.9        & 20.1  \\ 
\rowcolor{c3}SPC & Wide FOV    & 4    & 825        & 11.40\%  & 5.9        & 20.1 \\
\hline
\end{tabular}
\caption{Bandpass and coronagraph parameters are shown for the best effort polarimetric modes of Roman Coronagraph}
\label{tab:coronagraph-modes}
\end{center}
\end{table}

Although the requirement for the coronagraph is limited to a demonstration of TTR5 and is not constrained by any specific science requirement, previous studies conducted prior to the start of the Roman CPP program \cite{savransky2024nancy} demonstrated simulated polarization observations for both HLC and SPC modes. These studies also verified a science requirement established during the early design phase: to “map the linear polarization of a circumstellar debris disk that has a polarization fraction greater than or equal to 0.3 with an uncertainty of less than 0.03” \cite{anche2022simulations, anche2023simulation, anche2024high}. With the development of a high-fidelity end-to-end simulation package called \href{https://github.com/roman-corgi/corgisim}{\texttt{corgisim}}\cite{millar2024roman,zhang2026}, simulations of mock observation datasets can now be generated using the estimated observing parameters and processed through the data reduction pipeline \href{https://github.com/roman-corgi/corgidrp}{\texttt{corgidrp}}\cite{wang2026} to support the planning and execution of observations. Furthermore, the reduced datasets from \href{https://github.com/roman-corgi/corgidrp}{\texttt{corgidrp}} may facilitate the development and refinement of postprocessing algorithms, particularly for circumstellar debris disks.

This paper presents an overview of the polarimetric mode, including a description of the polarimetric observing scenario, exposure time estimation, and a simulation of an on-sky dataset for a representative observation of the circumstellar debris disk HD172555 using the HLC mode \cite{wolff2026, wolff2024roman}. The data reduction process is carried out with \href{https://github.com/roman-corgi/corgidrp}{\texttt{corgidrp}} to obtain the final Stokes parameters. Additional contributions at this conference address the calibration of the polarimetric mode \cite{mizuki2026}, observation planning for all campaigns, precursor observations for reference star vetting, features of \href{https://github.com/roman-corgi/corgisim}{\texttt{corgisim}}, and the architecture and development of \href{https://github.com/roman-corgi/corgidrp}{\texttt{corgidrp}}.

Section \ref{sec:pol_obs} details the polarization observing scenario and estimation of exposure time for polarimetric observations. Section \ref{sec:corgisim} presents the simulation of on-sky datasets using \texttt{corgisim}, while Section \ref{sec:corgidrp} describes the data reduction process using \texttt{corgidrp}. Conclusions and discussions are summarized in Section \ref{sec:discussion}.

\section{Polarimetric observing sequence}
\label{sec:pol_obs}
\begin{figure}[!ht]
    \centering
    \includegraphics[width=1\linewidth]{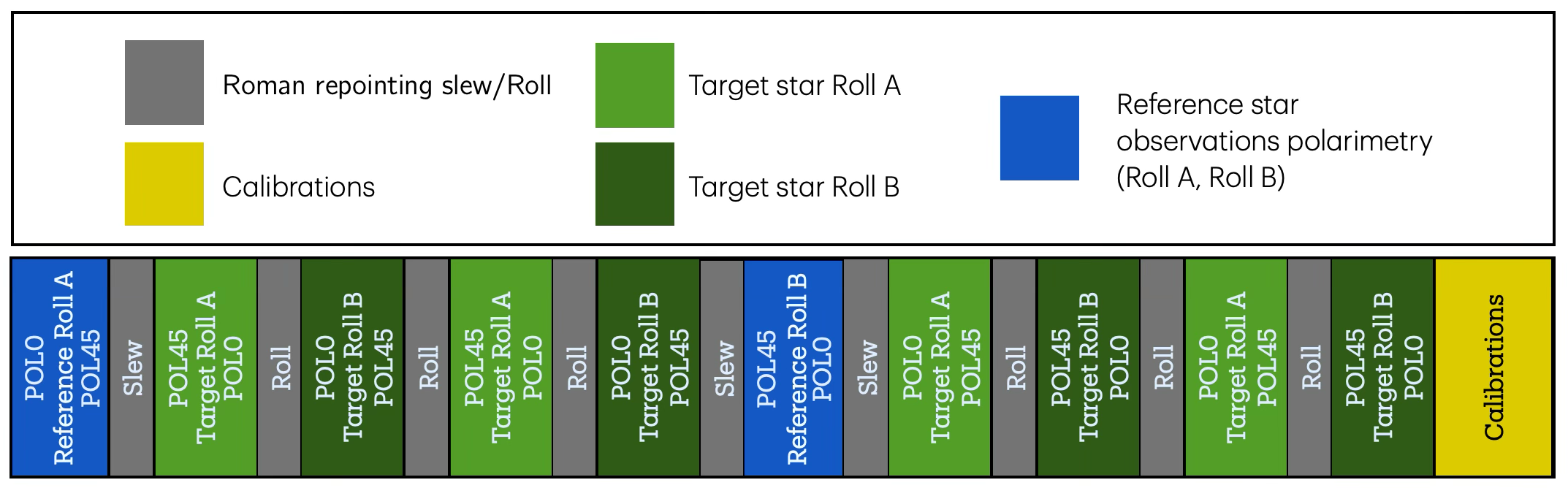}
    \caption{Polarization observation sequence in a observing campaign. The reference and target star are both observed in POL0 and POL45 wollastons at every roll angle. At end of the observation sequence, polarimetric calibrations are obtained}
    \label{fig:os_scenario}
\end{figure}

The polarization observation sequence illustrated in Figure \ref{fig:os_scenario} for an observing campaign involving multiple reference and target star visits begins with the reference star observed in both Wollaston configurations at the Roll A position, followed by slewing to the target star. Observations in both Wollaston configurations are then acquired for alternate roll positions on the target star, after which the reference star is observed at Roll B. Additional alternate roll positions for the target star may be included as necessary, and calibration observations are collected at the end of the sequence. During each reference and target star visit, satellite spots are acquired in both POL0 and POL45 at every roll position to help align individual frames. Reference star observations are obtained in both Wollaston configurations to enable post-processing using reference differential imaging (RDI) or polarimetric-constrained reference star differential imaging (PCRDI).
\subsection{Polarimetric calibrations}
Four distinct calibrations are necessary to address uncertainties specific to the polarimetric mode that are not resolved by standard imaging calibrations.
\begin{itemize}
    \item \textbf{Polarimetric flatfield}: This calibration corrects spatial variations in relative throughput and detector gain across the focal plane for images acquired with POL0 and POL45. Polarimetric flatfields are obtained using the same procedure as imaging flatfields, but with POL0 and POL45 positioned in the DPAM. Uranus or Neptune, depending on visibility, serve as sources for generating the flat fields \cite{maier2022flatfield}.
    
    \item \textbf{Mueller Matrix}: This calibration quantifies polarization effects introduced by the telescope and coronagraph instrument. The mirrors and optical components alter the polarization state of incoming light, which is characterized using a Mueller matrix. Observations of both polarized and unpolarized standard stars are required to estimate the instrument's Mueller matrix \cite{}. These observations are conducted in all configurations used for science data collection: with a neutral density filter, with various focal plane masks, and with different filters.

    \item \textbf{Polarization absolute flux}: This calibration converts detector digital numbers to physical flux units for a given spectral type using both Wollastons, POL0 and POL45. Bright and faint photometric standard stars are observed with the Wollastons for absolute flux calibration.

    \item \textbf{Reference and target polarimetry}: This calibration corrects for host star polarization by collecting unocculted reference and target star polarimetric data with a neutral density filter. The host star polarization is then subtracted from the polarization measured for the planetary companion or circumstellar debris disks.
\end{itemize}

\subsection{Exposure time estimates for polarimetric observations}
RDI scenario SNR equation and time to reach SNR is given in Nemati et al 2023 as: \cite{nemati2023analytical} 
\begin{eqnarray*}
SNR &= &\frac{r_{pl}t}{\sqrt{r_{n}t+f_{\Delta I}^2 r_{sp}^2 t^2}}\\
t_{SNR}&=&\frac{SNR^2 r_n} {r_{pl}^2 - SNR^2 f_{\Delta I}^2 r_{sp}^2 }
\end{eqnarray*}
$r_{pl}$ is the planet signal count rate and t is the integration time, $r_n$ is shot noise and detector noise contributions, and $f_{\Delta I}^2 r_{sp}^2$ arises from any lumpiness in the residual speckle image after differential imaging. 

\begin{figure}[!ht]
    \centering
    \includegraphics[width=0.4\linewidth]{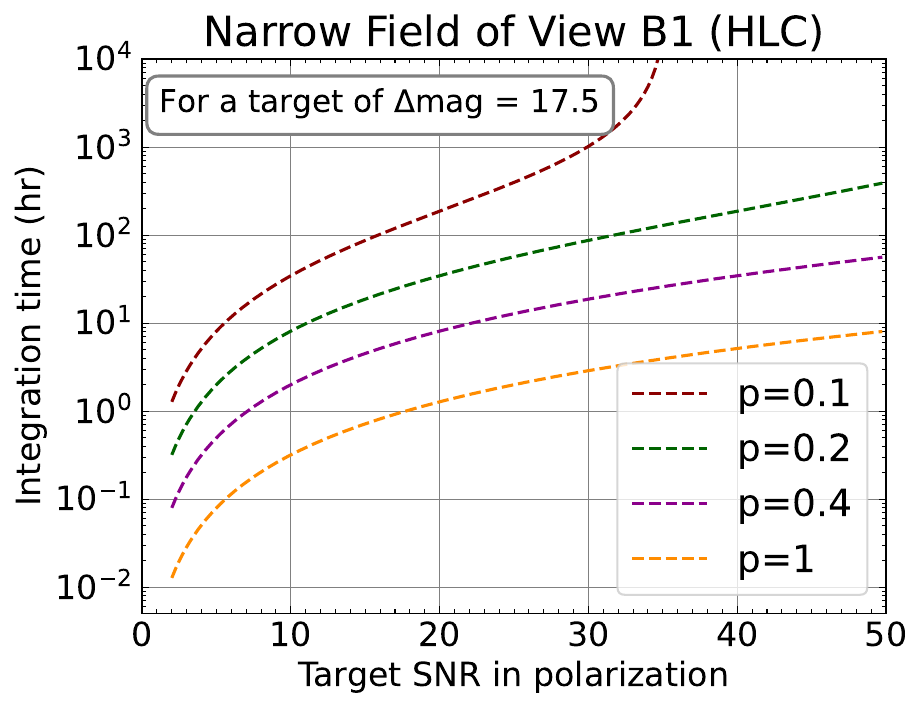}
     \includegraphics[width=0.4\linewidth]{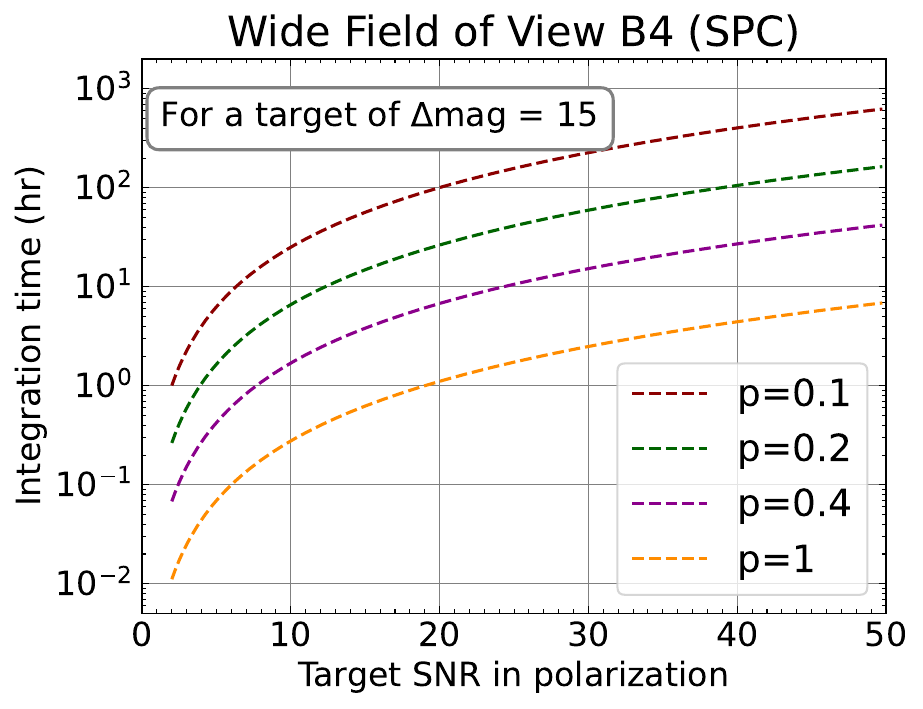}
    \caption{Exposure time estimates vs target SNR is shown for $\Delta$ mag of 17.5 and 15 is shown for narrow field of view band 1 (left) and wide field of view band 4 (right) for different values of polarization fraction }
    \label{fig:exposure_times}
\end{figure}

Given a known polarization fraction ($p_{m}$) and position angle ($\theta_{m}$), the analysis accounts for transmission through the Wollaston prism and incorporates modeled instrument Mueller matrices. The input polarization fraction does not depend on exposure time; however, the signal-to-noise ratio (SNR) required to achieve a specific polarization fraction is a function of exposure time. The detector noise component in $r_n$ is doubled because measurements are performed on two apertures. Additionally, the polarization speckle contrast term $C_{pab}$ is included with the reference differential imaging (RDI) post-processing factor.
For a given polarization fraction $p$, the polarization signal-to-noise ratio ($SNR_{pol}$) is defined as $SNR_{pol} = SNR \times p$.

\begin{eqnarray*}
    SNR &= &\frac{SNR_{pol}}{p} \\
    \frac{SNR_{pol}}{p}&=&\frac{r_{pl}t}{\sqrt{C_b+(f_{\Delta I}^2 r_{sp}^2 t^2})}\\
SNR_{pol}&=&\frac{r_{pl}^2t^2 p^2}{r_npol~t+f_{\Delta I}^2 r_{sp}^2 t^2+C_{pab}^2 t^2}\\
    t&=&\frac{SNR_{pol}^2 r_npol } {rpl^2 p^2 - SNR_{pol}^2 (f_{\Delta I}^2 r_{sp}^2+C_{pab}^2)} \\
\end{eqnarray*}
The polarimetric mode has been incorporated into \href{https://github.com/roman-corgi/corgietc}{\texttt{corgietc}} to estimate integration times for various polarization fractions. Figure \ref{fig:exposure_times} presents the required integration times for a target with $V_{mag}$=4.25 and a companion with $\Delta$mag=17.5 in HLC mode and $\Delta$mag=15 in SPC mode.

\section{Simulation of Level 1 (L1) data products in {\texttt{corgisim}}}
\label{sec:corgisim}
\begin{figure}[!ht]
    \centering
    \includegraphics[width=0.9\linewidth]{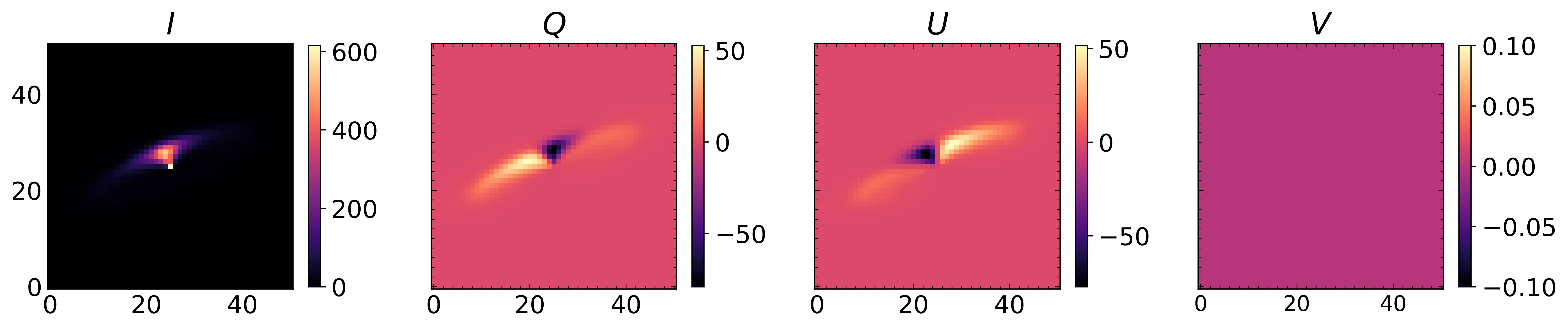}
    \caption{The input Stokes cube generated from MCFOST for debris disk around HD172555 for simulating through  \href{https://github.com/roman-corgi/corgisim}{\texttt{corgisim}}.}
    \label{fig:input_disk}
\end{figure}

In \href{https://github.com/roman-corgi/corgisim}{\texttt{corgisim}}, simulating a polarization observation of a planetary companion around a star requires providing a Stokes vector $[I, Q, U, V]$. In contrast, simulating polarization observations of circumstellar debris disks requires providing a two-dimensional Stokes cube.
The disk is initially modeled using the radiative transfer software \href{https://ipag.osug.fr/~pintec/mcfost/docs/html/overview.html}{MCFOST} \cite{pinte2006monte, pinte2009benchmark} to obtain the total intensity image and the Stokes parameter images $Q$ and $U$ for linear polarization. These outputs are subsequently used to generate a Stokes cube with dimensions $4\times n \times n$ and provided as an input to \texttt{corgisim}, as illustrated in Figure \ref{fig:input_disk} for the circumstellar debris disk around HD172555.
\begin{figure}[!ht]
    \centering
    \includegraphics[width=0.85\linewidth]{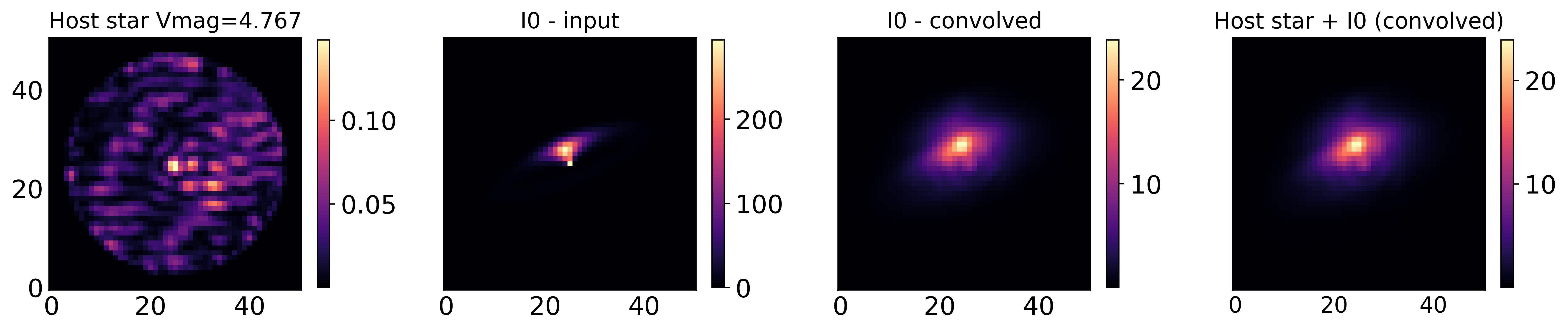}
    \includegraphics[width=0.85\linewidth]{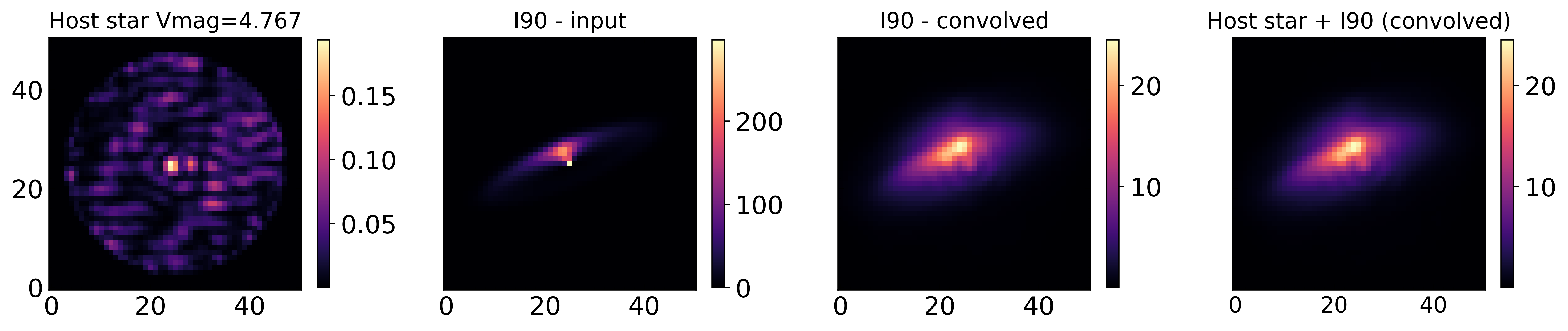}
    \caption{Polarization observations of HD172555 though POL0 wollaston prism showing $I_0$ and $I_{90}$ - host star and disk observations}
    \label{fig:convolved_disks}
\end{figure}
\begin{figure}[!ht]
    \centering
    \includegraphics[width=1\linewidth]{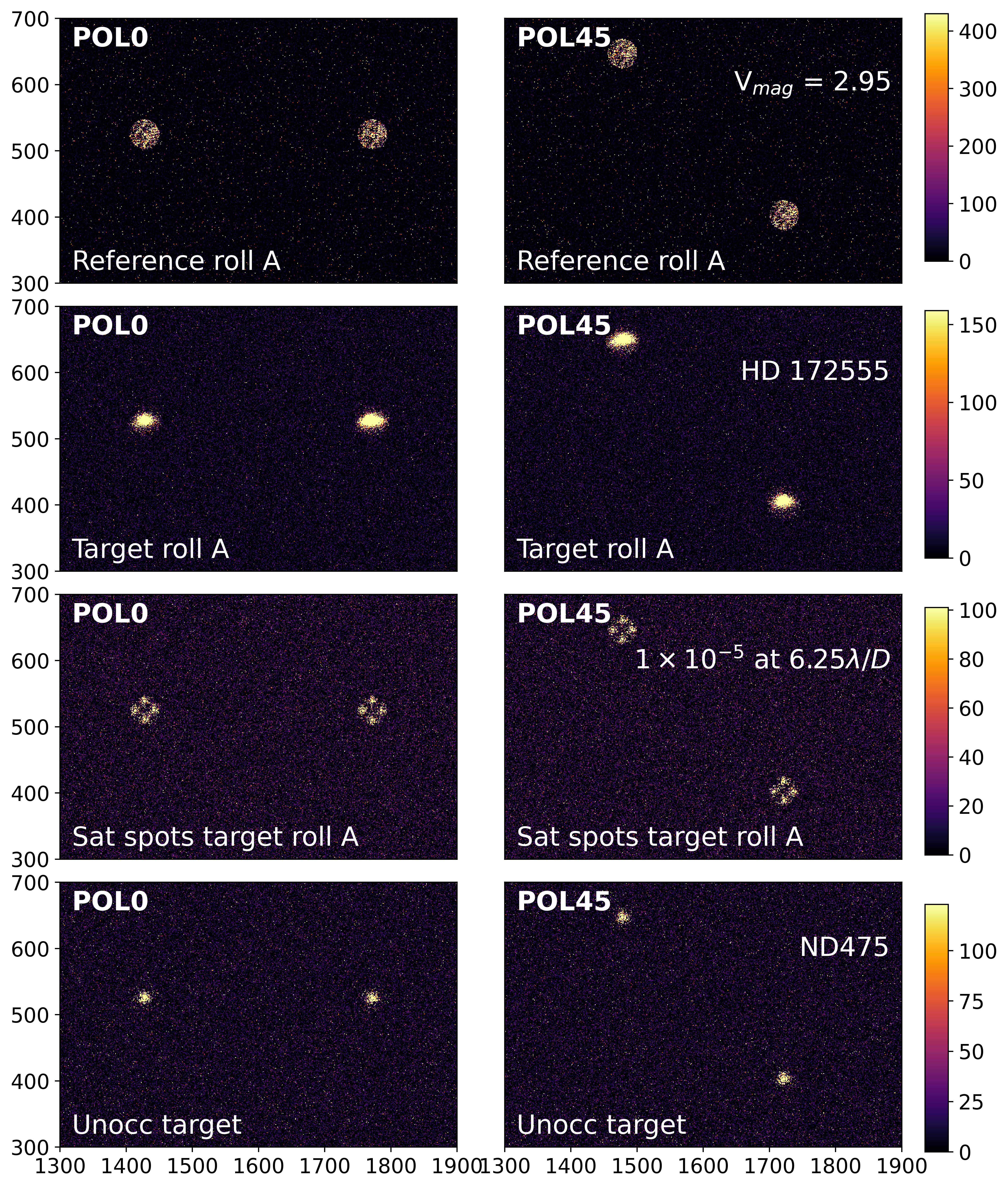}
    \caption{One of the  L1 frames from the polarization observation sequence of reference star $\alpha$ Ara (first row), target star HD172555 (second row), DM satellite spots for target (third row) and unocculted target star (fourth row) is shown for both the Wollaston prisms for one roll position.}
    \label{fig:l1_frames}
\end{figure}

The input Stokes cube is initially scaled to reflect the host star's magnitude and disk contrast, followed by multiplication with the modeled instrument Mueller matrices to incorporate polarization effects introduced by the Roman Coronagraph instrument. The resulting data are then multiplied by the Mueller matrix of the Wollaston prism, with POL0 used to obtain $I_0$ and $I_{90}$, and POL45 used to obtain $I_{45}$ and $I_{135}$. These outputs are subsequently convolved with the instrument point response function. The methodology for generating these point response functions is detailed by Anche et al. (2023) \cite{anche2022simulations}. Figure \ref{fig:convolved_disks} presents a coronagraphic image of the host star HD172555, along with the input and convolved disk images for $I_0$ and $I_{90}$ obtained using the POL0 Wollaston configuration. Following convolution, Excam full frames are simulated for both POL0 and POL45 using the previously estimated gain and integration time values.

To simulate a polarimetric observing scenario of a circumstellar debris disk using the NFOV mode in band 1, HD172555 was selected as the target and $\alpha$ Ara ($V_{mag}=2.95$, B2V) as the reference star \cite{hom2026coronagraph}.
HD172555 ($V_{mag}=4.7$) is an A7V star in the Beta Pictoris moving group, has a bright warm debris disk inclined at 76 degrees \cite{samland2025minds,engler2018detection,flasseur2020paco}, and has previously been observed with SPHERE/ZIMPOL using the very Broadband filter ($\lambda_c = 735$ nm, $\Delta \lambda = 290$ nm). $\alpha$ Ara serves as a vetted reference star without a bright companion within 6 arcseconds and is within the 5 degree delta pitch restriction \cite{hom2026coronagraph} at the time of observation \cite{wolff2026}.  

In addition to coronagraphic observations of the reference and target stars described in the observing scenario, $10^{-5}$ DM satellite spots are simulated at a separation of 6.25 $\lambda /D$ for both reference and target observations at both roll positions in POL0 and POL45. Unocculted reference and target observations in POL0 and POL45 are also simulated to estimate stellar polarization for both stars. All datasets are generated as full-frame EXCAM L1 images with the required header keywords for processing with \texttt{corgidrp}.

\section{Polarimetric data reduction through \texttt{corgidrp}}
\label{sec:corgidrp}
The \href{https://github.com/roman-corgi/corgidrp}{\texttt{corgidrp}} pipeline reduces L1 data products from each observation sequence to Technology Demonstration Analysis (TDA) data products. The processing steps follow the sequence: $L1 \rightarrow L2a \rightarrow L2b \rightarrow L3 \rightarrow L4 \rightarrow TDA$, as detailed in Wang et al 2026.
This section outlines the steps unique to the polarimetric observation sequence for observing a companion or extended sources during the processing of L1 data to TDA. In this reduction, only science frames were generated using \href{https://github.com/roman-corgi/corgisim}{\texttt{corgisim}}, and either mock calibration products or calibration products from the Roman Coronagraph Thermal Vacuum Chamber (TVAC) tests were utilized.  
\begin{figure}[!ht]
    \centering
    \includegraphics[width=1\linewidth]{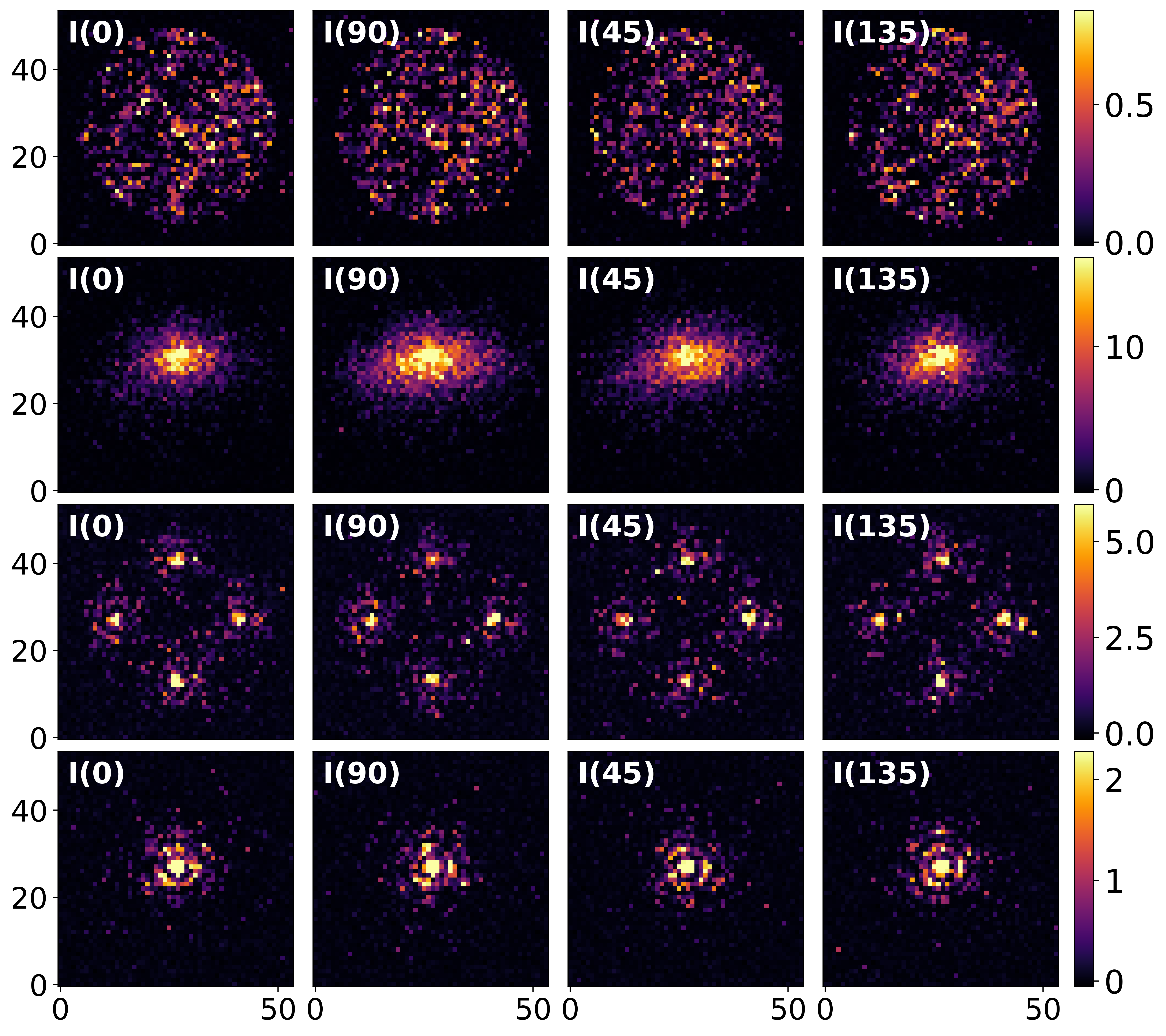}
    \caption{One of the  L3 frames from the polarization showing the extracted sub arrays for reference star $\alpha$ Ara (first row), target star HD172555 (second row), DM satellite spots for target (third row) and unocculted target star (fourth row) is shown for both the Wollaston prisms for one roll position.}
    \label{fig:l3_frames}
\end{figure}
\begin{figure}[!ht]
    \centering
    \includegraphics[width=1\linewidth]{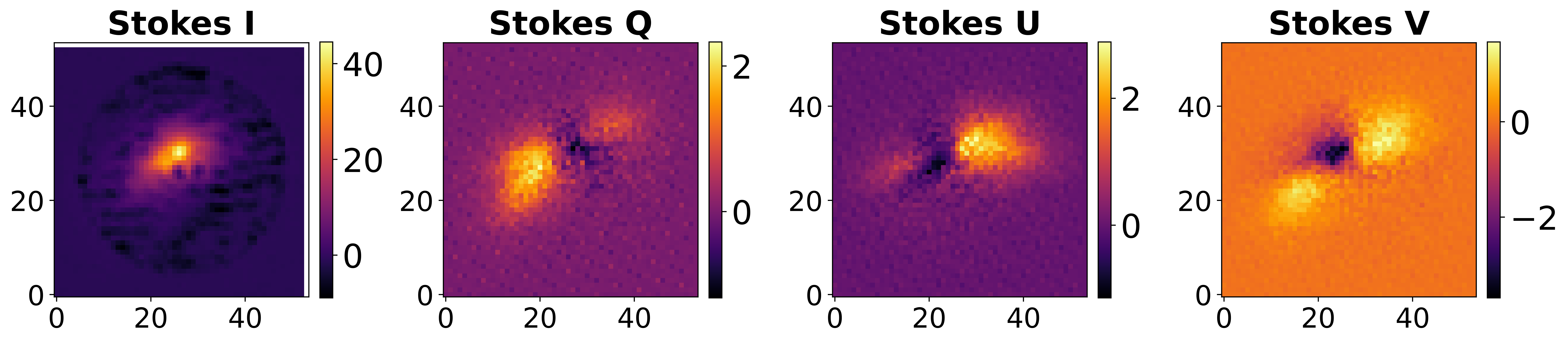}
    \caption{L4 level dataset showing Stokes cube with calibrated I, Q, U, and V}
    \label{fig:l4_frames}
\end{figure}
\begin{itemize}
   
   \item $L1 \rightarrow L2a$: Polarimetric images undergo the same processing steps as total intensity images, including bias subtraction, nonlinearity correction, and cosmic ray detection. L1 frames are full EMCCD frames with dimensions of 1200 by 2200 pixels, although the science data occupies only a small portion of the detector. Figure \ref{fig:l1_frames} presents example L1 frames focused on the science data region of the detector for observations using Wollaston, POL0, and POL45. These L1 frames are produced with all primary and extension headers, containing all keywords necessary for data reduction. In the observing scenario simulations conducted here, L1 frames are generated in the analog mode of the detector, with integration times estimated using the polarimetric mode in \texttt{corgietc}.
    
    \item $L2a \rightarrow L2b$: Polarimetric images are processed in the same manner as total intensity images, except that flat division is performed using a polarimetric flatfield. Additional steps include converting counts to electrons, dividing by electron multiplication gain, dark subtraction, charge transfer inefficiency correction, and correction of bad pixels. For this initial data reduction, a mock polarimetric flatfield, consisting of a 1024 $\times$ 1024 array of ones, is used. This flatfield will later be replaced with one generated from raster-scanned images of Uranus or Neptune obtained through POL0 and POL45.

   \item $L2b \rightarrow L3$: POL0 and POL45 images are separated into I(0), I(90), I(45), and I(135). Subarrays are extracted, and a single image cube is constructed, with slices corresponding to the polarization states for the specified Wollaston. The data are normalized by exposure time, and a World Coordinate System (WCS) is generated. Figure \ref{fig:l3_frames} presents the frames in L3 image cubes for observations through POL0 and POL45, including the reference star, target star, DM satellite spots for the target star, and the unocculted target star, all normalized by exposure time.

    \item $L3 \rightarrow L4$: All frames and subframes are aligned, and the host star polarization is subtracted from these frames. Unocculted observations of the reference and target stars are used to estimate stellar polarization. The Stokes cube, consisting of $[I, Q, U, V]$, is then estimated using I(0), I(90), I(45), and I(135). Frames are rotated to North up and East left, followed by point spread function (PSF) subtraction. Mueller matrix calibration is subsequently applied to estimate the calibrated Stokes cube. The Mueller matrix is generated using unocculted observations of polarized and unpolarized standards. In these simulations, a mock Mueller matrix calibration is used, which will be updated with Mueller matrix estimates from \href{https://github.com/roman-corgi/corgisim}{\texttt{corgisim}} simulations of selected polarized and unpolarized standards. Figure \ref{fig:l4_frames} presents the calibrated Stokes cube with I, Q, U, and V. Although the input Stokes cube simulated using MCFOST had zero circular polarization, nonzero $V$ is observed due to the modeled instrument Mueller matrices, which introduce crosstalk between $U$ and $V$.

   \item $L4 \rightarrow TDA$: The PSF-subtracted Stokes cube, calibrated for I, Q, U, and V, is used to estimate the following: a) for detected companions, polarization fraction, polarization signal-to-noise ratio (pSNR), and polarization position angle; b) for detected circumstellar disks, azimuthal Q and U components ($Q_\phi$ and $U_\phi$) and polarized intensity.

\end{itemize}

\section{Conclusions and Discussions}
\label{sec:discussion}

We present here the observing scenario, exposure time estimates, simulations, and data reduction for the polarimetric mode of the Roman Coronagraph instrument. 
The simulations utilized mock calibration products, specifically Mueller matrix calibration and flatfield data.
Prior to launch, simulations will incorporate calibration products generated with \href{https://github.com/roman-corgi/corgisim}{\texttt{corgisim}}. Complete observation sequences will be simulated using \href{https://github.com/roman-corgi/corgisim}{\texttt{corgisim}} and processed with \href{https://github.com/roman-corgi/corgidrp}{\texttt{corgidrp}}.

\acknowledgments 
Portions of this work were supported by the NASA under Grant 80NSSC25K0367  issued through the Roman Community Participation program (PI, Anche). J.N.A was supported by NASA through the NASA Hubble Fellowship grant \#HST-HF2-51547.001-A awarded by the Space Telescope Science Institute, which is operated by the Association of Universities for Research in Astronomy.
 Alexis Lau and Sophie Noiret acknowledge the support by the European Union (ERC, ESCAPE, project No. 101044152). Views and opinions expressed are, however, those of the author(s) only and do not necessarily reflect those of the European Union or the European Research Council Executive Agency.
The research was carried out in part at the Jet Propulsion Laboratory, California Institute of Technology, under a contract with the National Aeronautics and Space Administration (80NM0018D0004).

\bibliography{report} 
\bibliographystyle{spiebib} 

\end{document}